\documentclass[prologue, table, xcdraw]{vgtc}                 

\ifpdf
  \pdfoutput=1\relax                   
  \usepackage{graphicx}                
  \DeclareGraphicsExtensions{.pdf,.png,.jpg,.jpeg} 
\else
  \usepackage{graphicx}                
  \DeclareGraphicsExtensions{.eps}     
\fi%

\graphicspath{{figures/}{pictures/}{images/}{./}} 

\usepackage{microtype}                 
\PassOptionsToPackage{warn}{textcomp}  
\usepackage{textcomp}                  
\usepackage{mathptmx}                  
\usepackage{times}                     
\usepackage{cite}                      
\usepackage{tabu}                      
\usepackage{booktabs}                  

\usepackage{comment}
\usepackage{enumitem}
\newif\ifnotes
\notesfalse

\newcommand{\revised}[1]{\ifnotes{\leavevmode{\color{blue}#1}}\else{#1}\fi}

\onlineid{1097}

\vgtccategory{Research}

\vgtcinsertpkg

\title{Exploring D3 Implementation Challenges on Stack Overflow}

\author{Leilani Battle\thanks{e-mail: leibatt@cs.washington.edu}\\ %
        \scriptsize University of Washington %
\and Danni Feng\thanks{e-mail: fengdn@terpmail.umd.edu}\\ %
     \scriptsize University of Maryland %
\and Kelli Webber\thanks{e-mail: kwebber1@terpmail.umd.edu}\\ %
     \scriptsize University of Maryland}

\abstract{
Visualization languages help to standardize the process of designing effective visualizations, one of the most prominent being D3.
However, few researchers have analyzed at scale how users incorporate these languages into existing visualization programming processes, i.e., \emph{implementation workflows}.
In this paper, we present an analysis of the experiences of D3 users as observed through Stack Overflow, summarizing common D3 implementation workflows and challenges discussed online.
Our results show how the visualization community may be limiting its understanding of users' visualization implementation challenges by 
ignoring the larger context in which languages such as D3 are used.
Based on our findings, we suggest new research directions to enhance the user experience with visualization languages. All our data and code are available at: \revised{\url{https://osf.io/fup48/}}.
} 


\CCScatlist{
  \CCScatTwelve{Human-centered computing}{Visu\-al\-iza\-tion}{Visualization design and evaluation methods}{}
}

\begin{document}


\firstsection{Introduction}

\maketitle

\label{sec:introduction}
\vspace{-1mm}

Visualization languages provide great flexibility in programming and reusing visualization designs~\cite{bostock_d3_2011,mei2018design}.
Browser-based languages in particular have made it easier for a wide range of people to experiment with programming visualizations online~\cite{bostock_d3_2011,battle_beagle:_2018,satyanarayan_vega-lite:_2017}, one of the most influential being D3~\cite{bostock_d3_2011}. On the one hand, D3 is wildly successful: it has been starred over 101 thousand times on GitHub and has been identified on many thousands of webpages, including those of highly regarded media outlets~\cite{battle_beagle:_2018}.
On the other hand, D3 also has steep learning curve, and can be challenging for analysts and data enthusiasts to adopt~\cite{mei2018design,satyanarayan2019critical}.

One approach to addressing this problem is to investigate the challenges D3 users face with existing visualization programming processes and toolsets, i.e., \emph{implementation workflows}. This knowledge could be used to strengthen existing infrastructure for implementing D3 visualizations (e.g., documentation, examples, tutorials), and even lead to targeted support tools that integrate directly with these workflows. Furthermore, D3 is an important baseline for comparison with other visualization languages~\cite{satyanarayan_reactive_2016}, and existing languages adopt a similar support structure (e.g., Vega~\cite{satyanarayan_reactive_2016} and Vega-Lite~\cite{satyanarayan_vega-lite:_2017}). By investigating how to address implementation challenges for D3, similar improvements could be propagated to other languages.

However, we observe a dearth of academic corpora regarding how users interact with visualization languages such as D3.
Given the thousands of Stack Overflow posts made about D3, online forums could be a rich resource regarding how D3 users implement new visualizations, and the challenges they run into along the way.
Furthermore,
these forums could enable us to study D3 users on a scale not yet seen in existing evaluations~\cite{lam2012empirical}.

In this paper, we make a first step towards filling this gap by mining and analyzing 37,815 posts about D3 collected from Stack Overflow.
To ground our analysis, we evaluate how D3’s original design goals have been realized among these users
to improve compatibility, debugging, and performance~\cite{bostock_d3_2011}.
Through this work, we make the following contributions:
\begin{enumerate}[nosep]
    \item We present a mixed methods analysis of the experiences of D3 users as observed on Stack Overflow. To the best of our knowledge, this paper is the \emph{first} to center on online forums, which is known to be a critical resource for D3 users.
    \item Through our analysis, we provide: a collection of common implementation workflows discussed by D3 users on Stack Overflow, common challenges when integrating D3 into these workflows, and common debugging patterns for D3 programs.
    \item We contribute empirical support for several ``common knowledge'' assumptions made about D3 users that have yet to be rigorously evaluated by the visualization community, such as the frequent incorporation of non-visualization tools into D3 implementation workflows and persistent barriers to finding relevant D3 examples online.
    \item Based on our findings, we identify new areas for future visualization research. As one example, more research is needed on how to strategically design example galleries to boost user creativity and innovation, which would benefit not only D3 users but users of all visualization languages.
\end{enumerate}

\vspace{-1mm}
\section{Related Work}


\paragraph{Analyzing Online Forums.}
A number of projects investigate the practices~\cite{zhang_are_2018,mack2018characterizing,treude_how_2011}, attitudes~\cite{bosu_building_2013,kauer2021public} and impacts~\cite{jones2019rscience,ravi2014great} of engagement in various online Q\&A communities.
Others consider what impact online forums can have on individual users~\cite{anderson_discovering_2012,merchant_signals_2019,ardichvili_cultural_2006,ford_paradise_2016,slag2015one}. 
We utilize prior work to explore the intersection between analysis of online Q\&A communities and visualization implementation workflows and behaviors. We observe similar phenomena as those reported in prior work, such as notable proportion of unanswered questions~\cite{slag2015one}, strong associations between certain tools and libraries~\cite{chen2016mining}, prevalent code re-use~\cite{abdalkareem2017code}, and an emphasis on web development~\cite{barua2014developers}.
However, given our focus on visualization, we can speak to impacts on D3 users specifically, and adapt existing guidance to the direct interests and needs of the visualization community.  Furthermore, we believe this paper is the \emph{first} to analyze users' D3 implementation workflows by mining posts from online forums.

\paragraph{Visualization Mining, Modeling, and Analysis.}
Several projects highlight the value in analyzing existing designs from the web, such as website designs \cite{kumar_webzeitgeist:_2013}, visualization designs \cite{saleh_learning_2015,benson_end-users_2014,battle_beagle:_2018,hu_vizml:_2019,harper_converting_2018,harper_deconstructing_2014}, and even code structure~\cite{head_interactive_2018}. Online platforms such as bl.ocks.org~\cite{blocks} and Observable~\cite{observablehq} also help communities engage with visualizations online. A number of works analyze visualizations as images to automatically extract information about the author's implementation decisions  (e.g., \cite{poco_reverse-engineering_2017,savva_revision:_2011,jung_chartsense:_2017,battle_beagle:_2018,harper_deconstructing_2014,harper_converting_2018}). We extend prior work with new ways of automating the process of analyzing users' challenges in implementing visualizations, rather than just analyzing the visualizations themselves.
%
%
Our findings could guide the design of automated techniques, e.g., by extending systems such as VizML~\cite{hu_vizml:_2019} and Draco~\cite{moritz2018formalizing}, to identify and fill specific blind spots in a user's visualization knowledge in a way that aligns with existing user implementation workflows.

\vspace{-1mm}
\section{Analysis Overview}
\label{sec:research-questions}
\vspace{-1.5mm}

Our primary objective in this work is to identify opportunities to enhance the D3 visualization implementation process. To this end, we use the original D3 design goals to drive our analysis (compatibility, debugging, performance)~\cite{bostock_d3_2011}. For sake of space, we focus on two of these goals in this paper: \emph{compatibility} and \emph{debugging}\footnote{\revised{Our analysis of \emph{performance} is shared on OSF at the following wiki page: \url{https://osf.io/fup48/wiki/performance-analysis/}}.}.
We summarize each goal as a concrete question for our analysis:
\begin{enumerate}[nosep]
\item{\textbf{Compatibility}: How is D3 used in conjunction with other tools and environments?}
\item{\textbf{Debugging}: How do users explore, interpret, and augment the behavior of D3 code?}
\end{enumerate}

To answer these questions,
we downloaded relevant Stack Overflow posts,
using the Selenium~\cite{seleniumhq} and Beautiful Soup~\cite{beautifulsoup} Python libraries. We used 
using Stack Overflow's tag search to limit the corpus to posts that include the ``d3.js'' tag. \revised{Only posts that were still accessible on Stack Overflow at the time of analysis were included.}
With our approach, we are able to analyze \textbf{37,815 total StackOverflow posts} from 17,591 unique D3 users, showing the power of scale afforded by our techniques.

We apply a mixed methods analysis approach:
\begin{enumerate}[nosep]
\item{\textbf{Explore} a randomized sample of posts to identify patterns of potential interest as qualitative codes;}
\item{\textbf{Filter} the full 37,815 corpus using specific keyphrases derived from our qualitative codes and associated quotes;}
\item{\textbf{Count} observations of keyphrases on the full corpus to produce quantitative measures;}
\item{\textbf{Compare} our results to follow-up analyses of Stack Overflow posts, the D3 documentation, or relevant GitHub issues and release notes (if applicable), to provide additional context.}
\end{enumerate}

\revised{For example, to analyze external tools and libraries using these steps, we: (1) \textbf{explore} our randomized sample and qualitatively code mentions of various tools, e.g., mentions of React.js are coded as ``react'' and Angular.js as ``angular''; (2) \textbf{filter} relevant Stack Overflow posts and GitHub issues programmatically, using keywords derived from our qualitative findings, e.g., ``reactjs,'' ``react.js,'' ``angular,'' etc.; (3) \textbf{count} the observations of each tool/library from the posts that pass our filters; and (4) \textbf{compare} our results between our coded sample and the distribution of tools detected in the full Stack Overflow corpus.\footnote{\revised{A spreadsheet of this analysis is provided in our supplemental materials, available here: \url{https://osf.io/t8s37/.}}}}

In line with prior qualitative studies (e.g., \cite{jones2019rscience, mack2018characterizing}), we created a representative sample of Stack Overflow posts for the \textbf{explore} phase of our analysis pipeline.
This sample contains 817 posts randomly sampled through the year 2020.
Three authors manually coded all 817 Stack Overflow posts using \emph{descriptive} deductive and inductive codes. Deductive codes were used for known categories in visualization\footnote{\revised{See the appendix for examples of deductive codes used for visualization types (\autoref{sec:appendix:a}) and interaction types (\autoref{sec:appendix:b}).}}, such as visualization types (e.g., \cite{battle_beagle:_2018}) and interaction types \revised{(e.g., \cite{brehmer2013multi})}. Inductive codes labeled self-reported visualization and/or user characteristics, such as labeling when bar charts, Adobe Illustrator, or SQL were mentioned. \revised{Codes were not exclusive, and could be applied in parallel. The resulting codes and their frequencies are reported in \autoref{tab:qualitative:codes}.}

\revised{We used a multi-phase coding process.
In the first phase, all three
coders independently coded the same 20 posts, discussed disagreements, and refined problematic codes. Then, this process was repeated two more times, each time with 20 new posts (60 posts total). In the second phase, the remaining posts were coded by the three coders over a 13 week period. All three coders still regularly reviewed and discussed the codebook\footnote{\revised{The complete codebook is available in our supplemental materials \url{https://osf.io/8r79d/}, which includes code definitions and examples.}}; new potential codes were carefully evaluated and discussed, and any agreed upon codes were added to the codebook.}

Note that for most of our qualitative findings from the \textbf{explore} phase, we verify our results quantitatively using the full corpus \revised{(in \autoref{sec:compatibility:toolsets}, \autoref{sec:debugging:methods}, and \autoref{sec:debugging:examples}).}

{
\begin{table*}
\footnotesize
\centering
\caption{\revised{The resulting codes from our qualitative analysis of 817 sampled Stack Overflow posts (total observations in parentheses)}. The specific codes applied to each post are shared in our supplemental materials, available here: \url{https://osf.io/3m97y/}.}
\vspace{-1mm}
\revised{
\begin{tabular}{ll}
\hline
\textbf{Category}      & \textbf{Codes}                                                                                                                                                                                                                                                       \\ \hline
Learning D3            & \begin{tabular}[c]{@{}l@{}}existing-public-examples (179), new-to-d3 (89), add-text-labels (51), search-difficulty (46), previous-post (34),\\ best-practices (31), d3-documentation (28), novice-programmer (12), why-d3 (7), guide (2)\end{tabular} \\
\rowcolor[HTML]{EFEFEF} 
Errors \& Behaviors    & \begin{tabular}[c]{@{}l@{}}observed-unexpected-behavior (337), desired-output (222), observed-error (87), procedure (73), issue-not-d3 (44),\\ error-hypothesis (43), experimentation (12), verifying-correctness (9)\end{tabular}                                    \\
APIs                   & other-apis-tools (191), d3-method-{[}name{]} (86), d3-version-issues (50), import-d3 (25), d3-plugin-{[}name{]} (11)                                                                                                                                                       \\
\rowcolor[HTML]{EFEFEF} 
Visualization Details. & \begin{tabular}[c]{@{}l@{}}vis-{[}type{]} (398), interaction-{[}type{]} (181), data-{[}type{]} (154), animation (37), output-{[}type{]} (44), multidimensional (16),\\ real-time (10), 3d-model (4)\end{tabular}                                                                                                \\ \hline
\end{tabular}
}
\label{tab:qualitative:codes}
\vspace{-3mm}
\end{table*}
}
\vspace{-2mm}
\section{Compatibility: Integrating D3 With Other Tools}
\label{sec:compatibility}
\vspace{-1.5mm}

In this section, we analyze mentions of external APIs and tools in our random sample, and summarize the types of issues that Stack Overflow users discuss online.

\vspace{-2mm}
\subsection{Analyzing Users' Visualization Toolsets}
\label{sec:compatibility:toolsets}
\vspace{-1.5mm}

23.4\% of all posts we qualitatively analyzed referenced at least one other external library or tool. We found 55 different languages, tools, and libraries mentioned in conjunction with D3.js. To see if these frequencies suggest a larger pattern, we compared with observations of the same tools in the full Stack Overflow corpus.
We found that the random sample and full corpus have very similar distributions, pointing to broader patterns in how D3 is being used in conjunction with other tools (please see our technical report for more details).
We highlight a few interesting observations here.

\vspace{-1.5mm}
\paragraph{\revised{D3 is Often Integrated Into Larger Web Applications}.}
\revised{For example, D3 users often use} React, Angular, or Vue to manage the front-end, and Python, Node, Ruby on Rails, Electron, or Java Spring to run the back-end.
\revised{34} of 55 (or 61.8\%) of the tools we observed are JavaScript Libraries, and React.js, Angular, dc.js, jQuery, and NVD3 are mentioned the most.
However, each tool covers a small fraction of posts. For example, React is mentioned in 5\% of coded posts.
Thus there are a few popular application structures, but Stack Overflow users vary widely in the libraries/tools they use with D3.

\vspace{-1.5mm}
\paragraph{D3 is Paired With Specialized Visualization Libraries.} Though D3 provides support for geographic maps and graphs, we find some users attempting to integrate D3 with specialized visualization libraries, such as leaflet, simplify, datamaps, and openweathermap for maps, and web cola for graphs. Greensock and Three.js, specialized libraries for 2D and 3D animation, were also mentioned.

\vspace{-1.5mm}
\paragraph{D3 is Used Outside of JavaScript.} D3 is not used solely with JavaScript. For example, some Stack Overflow users seek help using D3 in R and Jupyter Notebooks. Users mentioned three packages in particular for R: shiny, r2d3, and radialnetworkr. Users also mention other computational environments, such as PostgreSQL and SparQL. However, the overwhelming majority of Stack Overflow posts that we analyzed focused on a JavaScript programming context.

\vspace{-1.5mm}
\paragraph{Graphics Editors and Spreadsheets are Used for Prototyping.}
We also find some interesting implementation variants, such as using Microsoft Excel or Adobe Illustrator for brainstorming prior to implementation (e.g., posts D-73 and A-296 from our dataset files, respectively). Three graphics editors were mentioned on Stack Overflow (Illustrator, CorelDraw, and InkScape), suggesting that a number of users brainstorm visualizations in non-code environments as part of their implementation workflows, and prior to using D3.

\vspace{-2mm}
\subsection{Common Assumptions Clash With User Workflows}
\vspace{-1.5mm}

When investigating one of the more popular application structures (React components), we find two common integration challenges. First, React is known to have a steep learning curve, and Stack Overflow users often encounter challenges simply in getting React to work properly. For example, we find that many posters have more trouble understanding React than D3 itself (e.g., post B-502). 
The second theme we observe is a clash in functionality between React and D3. D3 was originally designed to manipulate the DOM directly~\cite{bostock_d3_2011}, with known issues for integrating D3 with other libraries that also modify the DOM such as React~\cite{hackernoon}.
We observed this integration challenge in our qualitative analysis; for example, one poster's solution to their D3 integration problem was to use React-focused utilities designed to integrate \revised{with} D3, rather than use D3 directly (post B\_C-63).
We also observed Stack Overflow users mentioning libraries that replace D3's DOM manipulation operations with those of another library, such as how ngx-charts uses Angular instead of D3 for rendering purposes (e.g., post D-38).
Some answers even suggested removing D3 entirely (e.g., post B-149).

We observe in our analysis that when incorporating a JavaScript library that manipulates the DOM, a Stack Overflow user tends to anchor their workflow on the use of this library, since the DOM is the physical structure of the webpage itself. By choosing to manipulate the DOM directly, earlier versions of D3 tried to act as the anchoring library. Thus in a way, the design of D3 assumed that D3 was the focal point of an implementation workflow, even though D3 is scoped primarily for data interaction and visualization, which represent only a fraction of a user's overall interface and webpage design.
However, in reviewing D3's release notes, we found that D3's structure has shifted significantly over time from being an anchoring language to a modular collection of data manipulation and visualization libraries, likely \revised{due in part} to these challenges. \revised{Consider this excerpt from the D3v4 release notes~\cite{d32016release}:}
\vspace{-1.5mm}
\begin{quote}
\revised{\emph{D3 is now modular, composed of many small libraries that you can also use independently. Each library has its own repo and release cycle for faster development. The modular approach also improves the process for custom bundles and plugins.}}

\revised{\emph{There are a lot of improvements in 4.0: there were about as many commits in 4.0 as in all prior versions of D3. Some changes make D3 easier to learn and use, such as immutable selections. But there are lots of new features, too!}}
\end{quote}
\vspace{-1.5mm}

D3's evolution in response to user \revised{(and developer)} implementation challenges provides critical context for how visualization languages and tools can be designed in the future.
For example, the visualization community often develops and evaluates new tools without considering how they will be incorporated into existing user workflows \revised{\cite{plaisant2004challenge,shneiderman2006strategies}}. This lack of broader awareness could lead to less functional tools, and ultimately lower adoption and impact for visualization work. 


\vspace{-2mm}
\subsection{Takeaways.}
\vspace{-1.5mm}

We see a wide range of libraries and tools used as part of larger visualization implementation workflows, from specialized visualization libraries to usage with non-JavaScript languages such as Python and R, and even graphics editors and spreadsheets.
We explore how D3 evolved in response to users' implementation challenges, such as D3 competing with other libraries to manipulate the DOM. These challenges may speak to broader limitations in how visualization research is conducted, in particular the perspective that visualization tools are the focal point of a user's visualization implementation workflow.
Our findings suggest that the visualization community may benefit from taking the goal of compatibility even further, for example by treating visualizations as just one component of a larger application that a user wants to create, rather than the main focus. This shift in perspective necessitates a change in how visualization tools are designed and evaluated. Long term case studies could be a useful starting point~\cite{shneiderman2006strategies}, as they enable researchers to observe how people use visualization tools in real world environments over time.

\vspace{-1mm}
\section{Debugging: Interpreting \& Applying D3 Concepts}
\label{sec:debugging}
\vspace{-1.5mm}

In this section, we qualitatively analyze the kinds of bugs that Stack Overflow users often run into with D3, and the different strategies they use to articulate and fix their D3 implementation bugs.

\vspace{-1mm}
\subsection{Analyzing Implementation \& Debugging Methods}
\label{sec:debugging:methods}
\vspace{-1mm}

\paragraph{Stack Overflow Users Wrestle With Odd D3 Behavior More Often than Explicit Errors.}
Users mention explicit compilation or runtime errors only 10.6\% of the time (87 out of 817 posts). Instead, Stack Overflow users' bugs tend to involve runnable code that exhibits unexpected or unwanted behaviors (337 out of 817 posts, or 41.2\%), especially unexpected rendering effects or unexpected interaction behaviors in the visualization output.

\begin{figure*}
  \centering
  \includegraphics[width=0.9\textwidth]{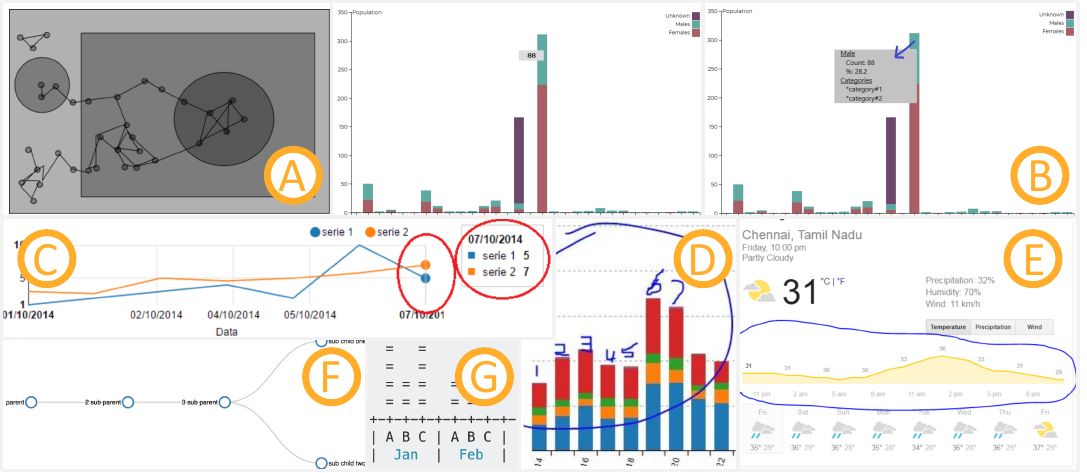}
  \vspace{-3mm}
  \caption{
  Example images shared online to convey desired visualizations (A, E, F, \&  G), interactions (B), and modifications (C \& D).
  }
  \label{fig:teaser}
  \vspace{-4mm}
\end{figure*}

\vspace{-1.5mm}
\paragraph{Stack Overflow users often rely on existing examples to find solutions to their bugs.}
We find that Stack Overflow users often reference existing D3 examples that are publicly available online when discussing their bugs (179 out of 817, or 21.9\% of posts). This suggests that Stack Overflow users often rely on existing examples when debugging their D3 code. To better understand the influence of the D3 documentation on example usage, we counted the number of Stack Overflow posts containing a link to examples from the visual index of the D3 visualization gallery, bl.ocks.org, and Observable~\cite{observablehq}. We find that 13.6\% of all posts directly reference examples from just these three sources, representing a significant fraction of all referenced examples in our qualitative dataset.

\revised{To} see if these findings point to a larger pattern, \revised{we} compare them with the full Stack Overflow corpus. Across all 37,815 posts analyzed, we find that 14\% include references to bl.ocks.org, Observable, or the D3 Gallery visual index, which is consistent with our coded data.
Thus \revised{D3} examples and documentation seem to be key components of users' D3 visualization implementation workflows.

\vspace{-2mm}
\subsection{Challenges in Using Relevant Examples to Fix Bugs}
\label{sec:debugging:examples}
\vspace{-1.5mm}

Adapting existing D3 examples seems to be an important but also complicated part of the D3 implementation and debugging process, which we explore further in this section.

\vspace{-1.5mm}
\paragraph{Some Users Struggle to Find the Most Relevant Examples.}
When searching for relevant examples, D3 users may struggle to match their own terminology for different visualization and interaction types to the terminology of others. For example, we find that 46 of 817 (or 5.6\%) of posts mention difficulties in finding \emph{any} relevant D3 examples or Stack Overflow posts.
If a D3 user is unaware of the appropriate D3 methods for implementing certain functionality, then a typical keyword search in Google or Observable may fail due to mismatches in terminology. Advanced visualization search engines may also be of limited use, due to their reliance on language-specific terminology, such as \cite{hoque2019searching} and \url{bl.ocksplorer.org}. However, once relevant examples are identified, D3 users still seem to struggle with separating relevant code from irrelevant code, and with distinguishing D3 components from those of other libraries.

\vspace{-1.5mm}
\paragraph{Complicated, Irrelevant Functionality Makes Examples Hard to Reuse.}
The utility of a D3 example seems to depend on not only the ratio of relevant to irrelevant functionality, but also the complexity of this functionality.
Unfortunately, existing examples routinely contain more functionality than the user wants or is familiar with, which can easily confuse the user, and lead to unnecessary bugs. Consider the following description from Stack Overflow, where a user is struggling to reuse the zoom functionality from a geographic map example that uses meshes:
\vspace{-2mm}
\begin{quote}
    ``\textit{I am trying to implement a 'zoom to bound' feature on my D3 map. This is the Block that I have in mind... My issue is that it looks like the implementation requires a topoJSON mesh.}'' (post A-587)
\end{quote}
\vspace{-2mm}
From follow-up comments on the post, we learn that meshes are not needed, revealing unnecessary functionality:
\vspace{-2mm}
\begin{quote}
    ``\textit{You don't need the mesh, that's just for the states' strokes.}'' (post A-587)
\end{quote}
\vspace{-2mm}
The commentor also suggests a more relevant example ``\textit{Have a look at this bl.ocks without mesh...}'' (post A-587), pointing to a separate problem that we observed: some users struggle to find the most relevant examples in the first place.

\vspace{-1mm}
\paragraph{Users Struggle to Connect D3 Code Components With Their Corresponding Visual Outputs.}
We find that many Stack Overflow users struggle to pinpoint the root causes of their D3 bugs.
One challenge for Stack Overflow users may be that the way in which users intuitively reason about visualization outputs may not match how these outputs are generally specified using the D3 language.
Rather than using the parlance of the D3 language to describe what they want to create, many Stack Overflow users illustrate their desired outputs by embedding one or more images within a post, often with annotations added. For example, 21.1\% of the posts we analyzed included linked or embedded images, which also held true for our full Stack Overflow corpus (19.2\%).

The annotations generally appeared to be hand-drawn, or applied using image editing or presentation software.
\autoref{fig:teaser} provides examples from five different Stack Overflow posts.
For example in \autoref{fig:teaser}-E, the user circled a specific visualization within an existing image (online weather information) that they want to copy.
In \autoref{fig:teaser}-G, the user has made a diagram in ASCII showing how they want to order and group the bars of a bar chart, reminiscent of ASCII-based diagramming seen in other debugging contexts (e.g., database debugging \cite{battle_making_2016}).
Some users would even go so far as to draw the specific changes they expect to make.
In \autoref{fig:teaser}-D, the user has manually illustrated their desired results by annotating a stacked bar chart with the totals drawn above each stack.
Users also illustrate desired interactions by sharing videos (e.g., post E-110), .gifs (e.g., post D-38), and sequences of annotated images (e.g., \autoref{fig:teaser}-A and B).

\vspace{-1mm}
\subsection{Takeaways.}
\vspace{-1mm}

We find that Stack Overflow users often post about unexpected behaviors rather than explicit errors.
We observe that one common implementation and debugging strategy among Stack Overflow users is to compare their code with relevant examples that are publicly available online.
However, Stack Overflow users
often struggle to find the most relevant D3 examples to inform their implementation and debugging process.
When they do find relevant examples, they may still struggle to extract the most relevant functionality.
This is due in part to how many D3 examples lack a modular structure, and often contain functionality that other users do not need, which can confuse these users and lead to unnecessary bugs.


If users could search for desired functionality without domain-specific keywords, they could find relevant D3 examples with less time and effort.
Furthermore, some users have found their own way of communicating desired visualization outputs through annotated images, videos, and gifs.
If Stack Overflow users could demonstrate their desired changes by \emph{directly manipulating the visualization output} (e.g., \cite{lee2015sketchinsight,saket_evaluating_2018,zong2020lyra2}), then they could debug their code using a search-by-example process, rather than a language or keyword-based search.
Then, visualization tools could compute the corresponding implementation deltas and update the code automatically. Debugging via direct manipulation has been proposed in other contexts, such as for updating Jupyter notebooks~\cite{wu2020b2} and debugging SQL queries~\cite{gathani2020debugging}, and could be adapted for visualization languages.
\vspace{-1.5mm}
\section{Discussion: Implications for Future Research}
\vspace{-1mm}

D3 has made incredible contributions to the visualization community. In this paper, we investigate opportunities to further enhance the experience of D3 users and of visualization language users in general.
We present an analysis of 37,815 posts made by D3 users on Stack Overflow.
We evaluate D3 from two perspectives: compatibility and debugging.
Our findings show that when we focus on developing visualization languages but not how and where people actually use them, we may struggle to fully understand the user experience, and may thus fail to fully identify and address the needs of these users.
By being mindful of how users interact with visualization languages and relay their implementation challenges, we can develop innovative strategies to enhance users' implementation processes, increase users' information access, and empower users to explore a wider range of effective visualization designs.
In this section, we highlight opportunities for future work based on our findings.

\vspace{-1.5mm}
\subsection{Emphasize Integration With Non-Visualization Tools}
\vspace{-1.5mm}

When visualizations become \revised{the} sole focus of visualization tool development, developers and researchers may build tools that conflict with other critical needs within users' implementation workflows, potentially hindering adoption.
As demonstrated through the evolution of D3, our community needs to shift its mindset towards building modular visualization \emph{components} that can integrate smoothly with other tools.
Furthermore, we encourage more formal evaluations of how users integrate new tools and languages (or not) into their implementation workflows over time~\cite{plaisant2004challenge,shneiderman2006strategies}.
For example, we encourage our community to conduct more large scale, quantitative studies of how visualization languages are used in popular development environments, e.g., Jupyter, Observable, and R Studio.

\vspace{-1.5mm}
\subsection{Better Support Infrastructure Could Boost Adoption}
\vspace{-1.5mm}

We find that users on Stack Overflow often rely on examples to implement and debug new D3 visualizations, but struggle to find relevant examples and
reuse them correctly.
Our findings point to two challenges in supporting current debugging workflows. First, we lack \emph{modularized building blocks} for implementing new visualizations in D3.
This issue may stem in part from D3's mixing of declarative specification of encodings with imperative specification of interactions, which is addressed in later languages such as Vega-Lite~\cite{satyanarayan_vega-lite:_2017}.
Second, the problem may not only be with code structure but also \emph{insufficient infrastructure} for helping users understand the flow of the code, i.e., how results propagate through the various parts of a D3 visualization. This issue has led to recent developments such as the Observable notebook environment~\cite{observablehq}, but Observable still expects users to manually segment their own code. Similar environments such as Jupyter notebooks suffer from the same problems. Both challenges highlight how D3 users struggle to break their implementation challenges down into modular, solvable pieces.

Based on our findings, we argue that both challenges could be addressed effectively by improving the \emph{support infrastructure} around D3, rather than by modifying D3 itself. For example,
we could develop more intuitive search interfaces that support search by example or search by demonstration, such as searching existing D3 examples for specific visual outputs or intended interaction behaviors. Furthermore, these issues could be addressed in a data-driven way by mining solutions directly from the thousands of existing D3 examples we observed.
For example, new tools could leverage this data to help users identify separate D3 components within existing examples, and extract only the components that are needed.
Development environments could also be augmented to automatically show relevant documentation, or recommend relevant code blocks, based on inferred user goals, experience levels, and potential biases.

\vspace{-1.5mm}
\subsection{Make Effective \revised{Example Gallery} Design Active Research}
\vspace{-1.5mm}

Our findings in \autoref{sec:debugging} show that a basic web search is just not good enough to help users find relevant D3 examples. Asking users to search for solutions using specialized D3 or visualization keywords are sub-optimal alternatives.
Even Stack Overflow is insufficient; translating a specific D3 bug into a self-contained Stack Overflow question requires significant time and effort~\cite{ford_paradise_2016}, and 37\% of the 37,815 posts analyzed were left unanswered by other users.
We believe these issues persist because our community views them as engineering rather than research problems.

In an effort to shift this perspective, we suggest some interesting research opportunities to expand existing design galleries. We could synthesize current best practices in visualization design as a diverse set of modularized visualization examples that all visualization language developers aim to provide.
Rather than expecting developers to create design galleries meeting these requirements, we could also find ways to automate the process of design gallery generation itself. This solution could involve a mixture of automation and the crowd, where automated processes are developed not only to detect gaps in existing galleries, but also to encourage users to fill these gaps with new examples. This approach could also be used to detect redundant or low quality examples and replace them automatically.
Developing design galleries and documentation takes time, whether for open-source languages such as Vega-Lite~\cite{satyanarayan_vega-lite:_2017}, or commercial APIs such as Plotly~\cite{plotly}. Automating the documentation process could speed up the learning process and dampen learning curves for users of \emph{all} visualization languages.

\vspace{-1.5mm}
\subsection{Takeaways Summary}
\label{sec:discussion:summary}
\vspace{-1.5mm}

Here we list three major takeaways derived from our research:
\begin{itemize}[nosep]
    \item Develop and test visualization languages as part of larger implementation workflows involving multiple tools.
    \item Provide support infrastructure for helping users find relevant examples, extract meaningful code components, and integrate these components into their workflows.
    \item Automate the example gallery generation process, to ease the difficulty of designing effective examples and make the process more consistent across languages/tools.
\end{itemize}


\vspace{-1.5mm}
\subsection{Limitations and Future Work}
\vspace{-1.5mm}

One limitation is that
we only focus on D3 users who post on Stack Overflow, a subset of all D3 users. However, we are still able to study 17,591 total D3 users, showing the scale afforded by our approach.
Given that posters do not have to share personal information on Stack Overflow, user characteristics are not consistently available in our dataset; thus we exclude them from our analysis.
An interesting direction for future research is to conduct follow up interviews with D3 users to better understand their backgrounds, motivations, and experiences, providing additional context for our findings.
However, we believe that our ability to analyze thousands of D3 users helps to balance this limitation out.
Certain D3 functionality may not be well represented in our dataset, e.g., animations. It would be interesting to introduce filters to our Stack Overflow crawler to extract specific posts for more targeted analyses in the future, e.g., downloading animation-focused posts for further analysis.
In general, we hope that by sharing our materials, we can empower the community to explore visualization languages in new ways. For example, a promising avenue of future work could be
to analyze iteration on visualization languages and user reasoning in tandem over time.

\acknowledgments{
This research was funded in part by NSF Award \#1850115. We thank Dominik Moritz, Zhicheng Liu, Niklas Elmqvist, the UMD Human-Computer Interaction Lab, the BAD Lab, the UW Interactive Data Lab, and our (many) reviewers for their invaluable feedback. We also thank Arjun Nair and Rishik Narayana for their help with data collection.}

\bibliographystyle{abbrv-doi}

\bibliography{references}

\appendix

\section{\revised{Observed Visualization Types}}
\label{sec:appendix:a}

\revised{In this appendix, we report on the distribution of visualization types observed in our sample of 817 Stack Overflow posts, and explore potential reasons for why we see this particular distribution of visualization types in our sample.}

\paragraph{\revised{Coding Method.}}
\revised{We use the visualization taxonomy proposed by Battle et al.~\cite{battle_beagle:_2018} as a starting point for establishing deductive codes for visualization types, since this taxonomy is based on visualizations shared by users of D3, as well as other tools such as Plotly and Fusion Charts. Battle et al. observed 24 visualization types:
\texttt{area, bar, box, bubble, chord, contour, donut, filled-line
geographic map, graph/tree, heatmap, hexabin, line, radial/radar, pie,
sankey, scatter, treemap, voronoi, waffle, word cloud,
sunburst, stream graph}, and \texttt{parallel coordinates.}
We extended this taxonomy with nine more visualization types observed on Stack Overflow (\texttt{bullet, funnel, hive, marimekko, OHLC, polygon, table, gantt}, and \texttt{waterfall}), and three more types from the D3 gallery visual index (\texttt{dial, icicle}, and \texttt{horizon}). The final taxonomy includes $24+8+4=36$ visualization types.}

\begin{figure}
\centering
\includegraphics[width=1.0\columnwidth]{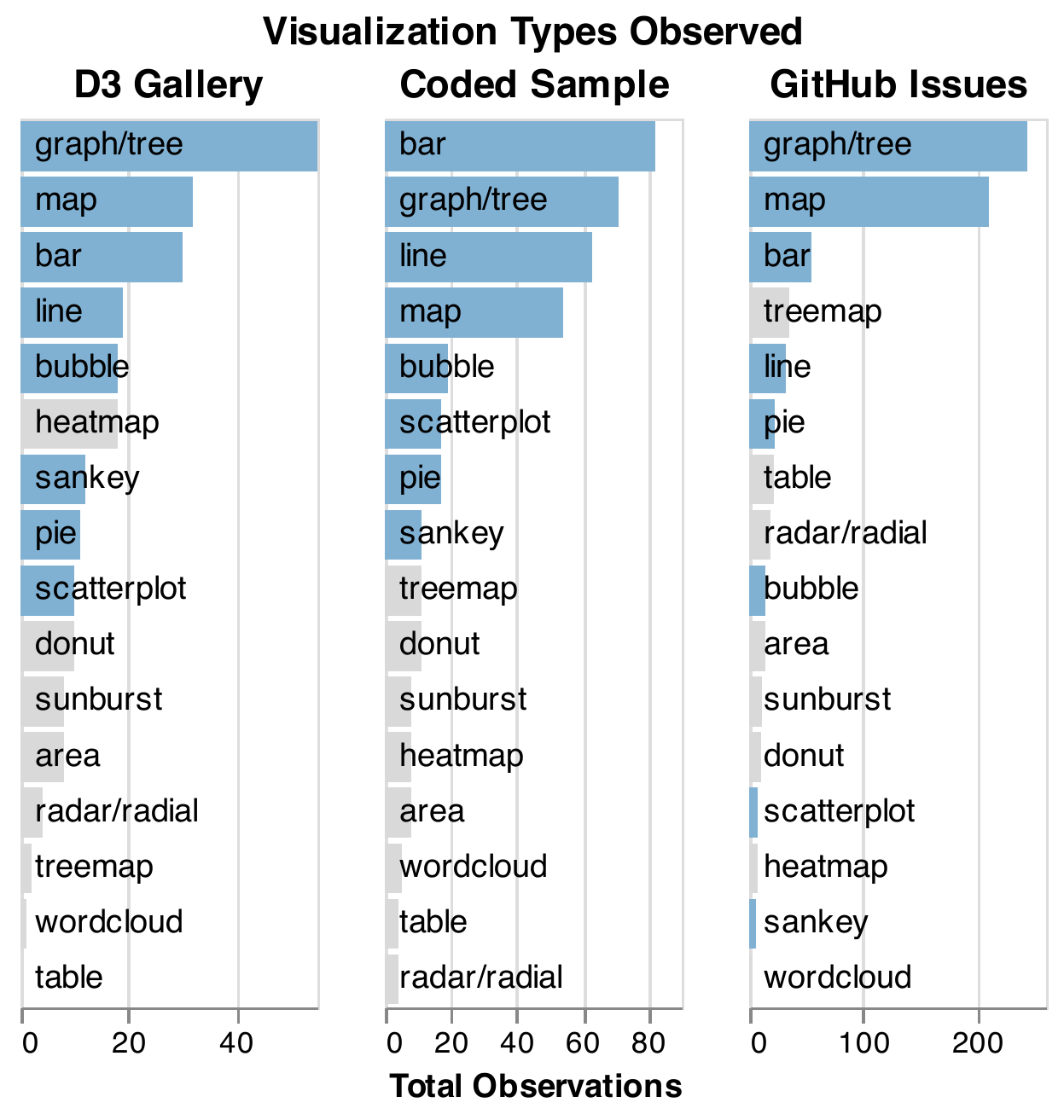}
\vspace{-5mm}
\caption{
\revised{Total observations for the top 16 visualization types observed in (a) the D3 gallery, (b) our randomized sample of qualitatively coded Stack Overflow posts, and (c) GitHub issues for D3. The top eight visualizations from the coded sample are colored blue.}}
\label{fig:all-vis-types}
\vspace{-6mm}
\end{figure}

\revised{We labeled each Stack Overflow post with any mentioned visualization types from our extended taxonomy. For example, when a user mentions bar charts: ``I'm working on creating a stacked bar chart...'' (post A-207) , we label the post with ``vis-bar''. Then, we tabulate the visualization types observed across the 817-post sample. We repeated this same coding process for issues discussed on the D3 github page, as well as the public gallery of visualizations shared on the D3 website. The top 16 visualization types observed for each corpus are reported in \autoref{fig:all-vis-types}.}

\paragraph{\revised{Stack Overflow Users Favor A Handful of Visualization Types.}}
\revised{We find that most posts focus on a specific visualization type.
We find that 48.7\% of our coded posts mention explicit visualization types; the top 16 are shown in \autoref{fig:all-vis-types} (center).
29 of the 36 visualization types from our extended taxonomy were observed (or 80.6\%), however 12 appear only once (or 33.3\%). One might assume that the most common visualization types (e.g., \texttt{bar} charts and \texttt{line} charts) are least likely to cause problems for Stack Overflow users, given their pervasiveness and simplicity.
However, the top three visualizations we observed were \texttt{bar} charts, \texttt{line} charts, and geographic \texttt{map}s, which are also three of the most common visualization types observed by Battle et al.~\cite{battle_beagle:_2018}.
More complex visualizations, such as \texttt{parallel coordinates}, are rarely mentioned in comparison. These results suggest that observed visualization types on Stack Overflow are indicative of user visualization preferences rather than complexity or challenge in implementing specific visualization types.}

\paragraph{\revised{Observed Visualizations on Stack Overflow Match the D3 Gallery.}} \revised{We compared the occurrence of visualization types in our coded dataset to those in the D3 example gallery, shown in \autoref{fig:all-vis-types}.
30 of 36 visualization types were observed in the D3 gallery visual index (or 83\%). \texttt{graph}s (and \texttt{tree}s) are the most popular visualization type in D3 gallery visual index, along with \texttt{map}s, \texttt{bar} charts, and \texttt{line} charts. Furthermore, we see that the top eight visualization types mentioned in our coded sample closely match the top eight visualization types observed in the D3 gallery. In contrast, the top visualization types mentioned on github only match roughly half of our coded sample.}

\revised{To gauge the importance of the D3 example gallery within the documentation, we analyzed how often the gallery page was updated on GitHub compared to other pages, e.g., the home
~\cite{d3home} or API reference~\cite{d3api} pages.
At the time of our analysis, the D3 Gallery page had been updated 1,295 times, three times as often as any other main page in the D3 documentation. We also found 29 GitHub issues that cited the D3 gallery visual index, supporting our findings.}

\paragraph{\revised{Takeaways.}}
\revised{When we combine our findings across GitHub issues, the D3 gallery, and in previous work~\cite{battle_beagle:_2018}, our results paint a more holistic picture of users' design choices.
We see that \texttt{line} and \texttt{bar} visualizations may dominate because they are universally popular across visualization tools, as observed in prior work~\cite{battle_beagle:_2018}. \texttt{graph} and \texttt{tree} visualizations may be of interest to Stack Overflow users because they appear frequently in the D3 gallery, documentation, and GitHub issues.
These results may suggest that D3 users rely heavily on existing D3 examples and documentation when implementing new visualizations. Thus the D3 gallery likely represents, and perhaps even influences, the range of visualizations created by Stack Overflow users.
As a result, these findings may also suggest that if the D3 documentation is skewed to prioritize specific visualization types, it could potentially introduce bias in to users' implementation workflows.}

\revised{Given the breadth of visualization types observed in the academic literature, one interesting question for the future is why Stack Overflow users do not seem to experiment with these other visualization types. On the one hand, these findings might suggest that the full design breadth of D3 may be interesting but also overkill for most Stack Overflow users. On the other hand, given that Stack Overflow users frequently reference existing D3 visualizations as part of their implementation workflows, the issue could be that there is insufficient scaffolding for users to feel confident in creating these other types of visualizations. The quantity and variety of gallery examples may be just as important as the quality of examples in helping users expand their design thinking.}

\section{\revised{Observed Interaction Types}}
\label{sec:appendix:b}

\revised{In this appendix, we report on the distribution of interaction types observed in our sample of 817 Stack Overflow posts, and compare our findings with the corresponding descriptions of these interactions in the D3 documentation.}

\paragraph{\revised{Coding Method.}} \revised{We use the typology proposed by Brehmer and Munzner~\cite{brehmer2013multi} to establish deductive codes for interaction types. We focus on the ``manipulate'' interactions in the typology, which summarize and align closely with other interaction taxonomies for information visualization (e.g., \cite{yi2007toward}). Brehmer and Munzner define six ``manipulate'' interactions and a separate ``encode'' interaction, which we use in our analysis:}
\revised{
\begin{itemize}[nosep]
\item \texttt{encode}: change the encodings
\item \texttt{select}: hover, click, lasso, or otherwise highlight marks
\item \texttt{navigate}: pan, zoom, rotate
\item \texttt{arrange}: reorder axes, change spatial layout
\item \texttt{change}: alter/format the visualization (not the encodings)
\item \texttt{filter}: include/exclude data records
\item \texttt{aggregate}: group, adjust granularity.
\end{itemize}
}
\revised{We stress however that this is only a subset of the ``how'' level of the Brehmer and Munzner typology, and expert D3 implementationers likely implement the full range of interactions, which include not only ``encode'' and ``manipulate'' interactions but also ``introduce'' interactions (\texttt{annotate}, \texttt{import}, \texttt{derive}, \texttt{record}). We focus on ``encode'' and ``manipulate'' interactions from the typology because they are likely to be most familiar to Stack Overflow users, but we also mention coverage of ``introduce'' interactions in our analysis.
}

\begin{figure}
  \centering
  \includegraphics[width=0.6\columnwidth]{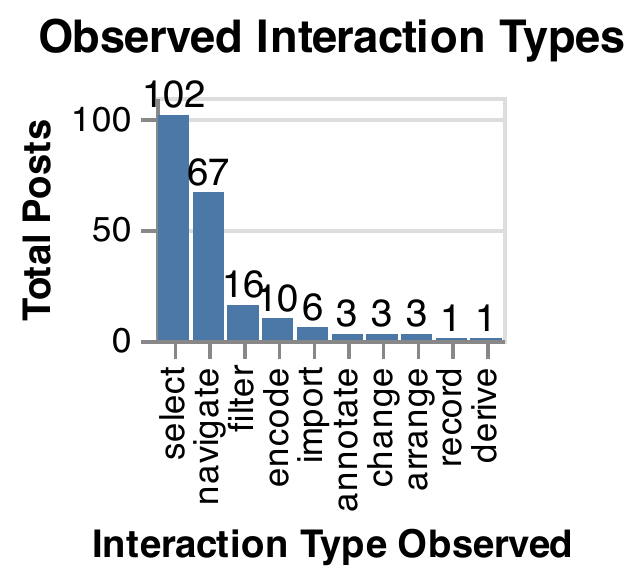}
  \vspace{-4mm}
  \caption{\revised{Total observations (y axis) for all interaction types from our taxonomy (x axis) in the qualitatively coded Stack Overflow data.}}
  \label{fig:interaction-types}
  \vspace{-6mm}
\end{figure}

\paragraph{\revised{Stack Overflow Users Favor a Few Interaction Types.}}
\revised{Our results are shown in \autoref{fig:interaction-types}.
We observed ``encode'' interactions, five of six ``manipulate'' interactions, and all four ``introduce'' interactions.
``Manipulate'' interactions (20.4\% of posts) were far more prevalent than ``encode'' (1.2\%) and ``introduce'' interactions (1.1\%). Given our results for visualization types, ``manipulate'' interactions are probably more popular and thus discussed more online. However, just two of the 11 interaction types (\texttt{select} and \texttt{navigate}) represent over 83\% of our observations.}

\revised{We see some Stack Overflow users combining interactions in their interfaces. Here is one example, where a user aims to display specific statistics on hover \texttt{selection}s, while also supporting \texttt{filter}ing:}
\vspace{-3mm}
\begin{quote}
    \revised{``\textit{... I want my charts to have the total Overall Packages number (Object) available so I can display that in the tooltip, etc. This object will not change during the filter, I just need that total.}'' (post A-528)}
\end{quote}
\vspace{-2.5mm}
\revised{However, Stack Overflow users tend to focus on one key interaction type in their code, often \texttt{select} or \texttt{navigate}.}

\paragraph{\revised{The Interactions in the D3 Documentation Match Observed Interactions on Stack Overflow.}}
\revised{To understand how the D3 documentation reflects the prevalence of certain interactions, we analyzed the the D3 API Reference. Three ``manipulate'' interactions seem to be well-represented in the documentation (with quoted topics in parentheses): \texttt{select} (``Search'', ``Brushes'', ``Selections''), \texttt{filter} (``Brushes'', ``selection.filter''), and \texttt{navigate} (``Dragging'', ``Zooming'').
Other interactions such as \texttt{change} have relevant topics (e.g., ``Time Formats''), but the documentation focuses on how to implement their functionality as immutable visualization components, rather than manipulable interaction widgets. These results suggest that the D3 documentation emphasizes a small subset of ``manipulate'' interactions, which correlates to a narrower range of interactions being discussed among Stack Overflow users.}

\paragraph{\revised{Takeaways.}} \revised{Not all users describe their challenges on Stack Overflow with the same quality and specificity~\cite{ravi2014great}. However, we do see some general trends.
For example, Stack Overflow users seem to favor a narrow subset of interaction types. We observe only five of seven ``manipulate'' interactions and one out of four ``introduce'' interactions from the Brehmer and Munzner typology. Subsequent analyses of the D3 documentation suggest that these interaction types are not described equally (if at all), suggesting that some interaction types are not prioritized among Stack Overflow users.
Taking these ideas a step further, the
D3 documentation could potentially be a \emph{driver} of this skew in user preferences, leading to visualization design bias among Stack Overflow users. However correlation does not equal causation; we leave quantitative evaluation of these hypotheses for future work.}

\end{document}